\RequirePackage{fix-cm}
\documentclass[twocolumn]{svjour3}          
\smartqed  
 \usepackage{amsmath,amssymb,latexsym,float,epsfig,subfigure,epstopdf,graphicx,array,algorithm,algpseudocode}
\usepackage[compress,square,comma,numbers]{natbib}

\RequirePackage[english=usenglishmax]{hyphsubst}
 \usepackage{mathptmx}      
\usepackage{latexsym}
\begingroup\emergencystretch=.5em
\begin{document}

\title{Finite Dictionary Variants of the Diffusion KLMS Algorithm
}


\author{Rangeet Mitra         \and
        Vimal Bhatia 
}


\institute{R. Mitra \at
              Signals and Software Group,               Discipline of Electrical Engg.,\\
              Indian Institute of Technology Indore \\
              Tel.: +91-8603011238\\
              \email{phd1301202010@iiti.ac.in}           
           \and
           V. Bhatia \at
           Associate Professor\\
           Signals and Software Group, Discipline of Electrical Engg.,\\ Indian Institute of Technology Indore \\
           Tel.: +91-07324240705\\
           \email{vbhatia@iiti.ac.in}
}

\date{Received: date / Accepted: date}

\maketitle

\begin{abstract}
	
The diffusion based distributed learning approach-es have been found to be a viable solution for learning over linearly separable datasets over a network. However, approaches till date are suitable for linearly separable datasets
and need to be extended to scenarios in which we need to learn a non-linearity. In such scenarios, the recently proposed diffusion kernel least mean squares (KLMS) has been found to be performing better than diffusion least mean squares (LMS). The drawback of diffusion KLMS is that it requires infinite storage for observations (also called dictionary). This paper formulates the diffusion KLMS in a fixed budget setting such that the storage requirement is curtailed while maintaining appreciable performance in terms of convergence. Simulations have been carried out to validate the two newly proposed algorithms named as quantised diffusion KLMS (QDKLMS) and fixed budget diffusion KLMS (FBDKLMS) against KLMS, which indicate that both the proposed algorithms deliver better performance as compared to the KLMS while reducing the dictionary size storage requirement.
\end{abstract}

\section{Introduction}
\label{intro}
In this age, we are flooded with huge volume of data based on which we aim to do inference. With the development of technologies \cite{boccardi2014five}, huge datasets distributed over networks have come into existence. There are two popular ways of processing this deluge of information, namely; a) Centralised processing, and b) Distributed processing. In centralised processing there is a central node which handles all the inference mechanism in the network via a central fusion center. This process has two major drawbacks: a) the central node has high computational requirement with scaling the network, and b) if the fusion center fails the entire centralised adaptation falls apart. To cater to these challenges, a new branch of adaptive filtering called diffusion distributed adaptive filtering \cite{cattivelli2008diffusion,cattivelli2011distributed,sayed2014adaptive}, has come into the scene of information processing. Examples of such scenarios occur in wireless sensor networks (see references in \cite{rastegarnia2015incremental}). The nodes rely on observations emanating from their neighbour nodes, hence reducing the per-node computational cost and at the same time delivering equivalent performance as compared to centralised based approaches.
\subsection{Related Works}
Many techniques have been proposed for distributed adaptive filtering in the existing literature. Incremental based approaches \cite{rastegarnia2015incremental,lopes2007incremental,sayeddistributed,ram2007stochastic,takahashi2008incremental,li2008new,li2010distributed,lopes2010randomized,
cattivelli2011analysis,saeed2012new,abadi2014low,bogdanovic2013distributed,liu2014enhanced,rastegarnia2013steady} visit each node cyclically and adapt the linear adaptive filter weights. However, it does not exploit all the data available to a particular node for inference. Diffusion based algorithms \cite{cattivelli2008diffusion,lopes2007diffusion,cattivelli2009multi,takahashi2009diffusion,cattivelli2010diffusion,gharehshiran2013distributed,
baqi2013diffusion,di2014distributed,chen2012diffusion,arablouei2014distributed} exploit the weighted local information from its neighbours and has been established as a viable solution for distributed adaptive filtering. This gives faster convergence as compared to incremental based approaches at the cost of slightly higher computational complexity.

Parallely, a new genre of adaptive filtering based on kernel adaptive filtering has become popular for non-linear inference. These algorithms are suited for scenarios when the incoming data is not linearly separable. Popular algorithms like least mean squares (LMS), affine projection algorithm (APA), recursive least squares have been mapped to reproducing kernel Hilbert space (RKHS) by the ``kernel trick". Most popular among them are kernel least mean squares (KLMS) \cite{liu2008kernel}, kernel recursive least squares (KRLS) \cite{engel2004kernel}, kernel affine projection algorithms (KAPA) \cite{liu2008kernelAPA}. These kernel adaptive filtering techniques theoretically require infinite memory requirements. To counter this shortcoming, algorithms like quantised KLMS and fixed budget quantised KLMS \cite{zhao2013fixed,chen2012quantized} have been proposed which adaptively restricts the size of the working dictionary.

The work in \cite{mitra2014diffusion}, called diffusion-KLMS combines the domains of kernel adaptive filtering with distributed diffusion adaptive filtering. However, like KLMS each node in the network is assumed to have infinite storage capacity.

\subsection{Motivation and Contributions}

In this paper, we address the important issue of restricting the infinite dictionary-size requirement for diffusion-KLMS, at all nodes in the network to a finite value. To this end, we propose two algorithms in this paper: a) the quantised diffusion KLMS (QDKLMS), which uses an online vector quantisation approach to update the distributed dictionary and b) fixed budget diffusion KLMS (FBDKLMS), which gives a way of pruning the dictionary over the network. These proposed approaches, QDKLMS and FBQDKLMS bounds the per-node storage requirement for diffusion-KLMS for estimating distributed non-linear hypotheses, and also paves the way for distributed non-linear hypothesis estimation in RKHS with a finite dictionary size for datasets in which a linear/affine separating boundary cannot be estimated.

\subsection{Paper Outline}
The outline of this paper is as follows: a) Sect. 2 reviews diffusion KLMS, b) Sect. 3 gives the QDKLMS, c) Sect. 4 gives FBDKLMS, d) simulations are provided to validate our proposed approaches in Sect. 5 and conclusions of this paper are drawn in Sect. 6.
	\section{Diffusion-KLMS}
	Based on the KLMS algorithm, we review its distributed variant in this section based on the diffusion approach. We now define matrices and symbols that will be used in this paper.
	In this proposal, we have the matrix $Y=[y_{l,n}]$ to denote output corresponding to the $l^{th}$ neighbour at $n^{th}$ time instant.
	$E=[e_{l,n}]$ is the error matrix corresponding to the $l^{th}$ neighbour at $n^{th}$ time instant.
	$X(n)=[\{\textbf{x}_{l,n}\}]$ is a matrix of measurement vectors  from neighbours of node $q$ at time instant $n$ stacked together.
	In the following few lines, we will denote the collection of the data from various nodes at the $n^{th}$ time instant as $X(n)$. $X(n)$ contains the data pertaining to all $l$ neighbours stacked in row vector form. In case, there is no vector from a node in the neighbourhood it is replaced by the zero vector in $X(n)$ and will have a corresponding 0 entry in ${C}$, which is a stochastic matrix of compatible dimension.
	
	Basically, the distributed version of diffusion KLMS can be formulated by considering the RBF analogy for KLMS in \cite{liu2008kernel}, with centers weighted by the innovations and considering it to be consisting of two steps namely the diffusion and incremental step:
	\begin{gather}
	y_{q,n}=\sum_{i=0}^{n-1}\delta_{q,i}<\phi(\textbf{x}_{i}),\phi(\textbf{x}_{q,n})>_{\mathcal{H}}
	\end{gather}
	where $<\cdot,\cdot>_{\mathcal{H}}$ denoting the kernel inner product in RKHS.
	The $\{\delta_{q,i}\}$ are weight-factor estimates at $q^{th}$ node at instant $i$ for RBF centers which happen to be the innovations $e_{q,i}$ at each node. Hence invoking the diffusion step (step-1) for the weighting factors similar to \cite{cattivelli2010diffusion}, we get:
	\begin{equation}
	e_{q,n}^{'}=\sum_{l}a_{ql}e_{l,n-1}
	\end{equation}
	Using these estimates of innovation-weights $\delta_{q,n}$ for a given node $q$ at the $n^{th}$ iteration, we use them for the incremental step (step-2) as follows:
	\begin{equation}
	\Omega_{q,n}=\sum_{i=0}^{n-1}\sum_{l \in \mathcal{N}_{q}}e_{q,i}^{'}c_{ql}<\phi(\textbf{x}_{l,i}),\cdot>_{\mathcal{H}}
	\end{equation}
	where $\Omega_{q,n}$ is the implicit learned parameter in RKHS.
	Applying the kernel trick results in,
	\begin{equation}\label{algo2}
	y_{q,n+1} = \eta\sum_{i=0}^{n-1} e_{q,n}^{'}<\phi(CX(i)),\phi(X(n))>_{\mathcal{H}}
	\end{equation}
$\eta$ being the step-size.
	\begin{equation}\label{algo1}
	e_{q,n+1}^{'} = \sum_{l\varepsilon \mathcal{N}_q} a_{ql} d_{l,n}-  \sum_{l \varepsilon \mathcal{N}_{q}} a_{ql} y_{l,n}
	\end{equation}
	where $A$ is a stochastic matrix corresponding to the probabilistic weights $\{a_{ql}\}$ and $\mathcal{N}_{q}$ denotes the node indices of neighbourhood of $q$.
	The error at $n^{th}$ time instant at the $q^{th}$ node would be the (transformed) mean (by A) of $e$ over all possible $l$.
	
	The proposed algorithm is given below, as iterating following three steps, till convergence:
	\begin{enumerate}
		\item Estimate the outputs of node $l$ using estimates of error $e_{l,n}^{'}$.
		\item Form an estimate of errors at time instant $n$ at each node $l$. Let this be given by the vector ${e}(n)$ whose $l^{th}$ element is $e_{l,n}$. Then the error term for the $l^{th}$ node for the $n^{th}$ time instant can be written as $e_{l,n} = d_{l,n} - y_{l,n}$, where $d_{l,n}$ is the desired value at $l^{th}$ node at $n^{th}$ time instant.
		\item The error at each node is modified by the transformation $A$ by the equation $\textbf{e}^{'}(n+1)=A\textbf{e}(n)$,
		where $\textbf{e}(n)$ and $\textbf{e}^{'}(n)$ are vectors of error terms corresponding to all the nodes (for all nodes indexed by $l$) stacked together.
	\end{enumerate}
\section{Proposed Algorithm I : Quantised Diffusion KLMS (QDKLMS)}
In this section, we propose a novel algorithm for distributed non-linear inference over a network with finite dictionary-size thereby lending it to efficient implementation.
The proposal presented in this section curtails the infinite arbitrary growth of the dictionary containing the innovations and observations.

Let us introduce the notion of a dynamic dictionary and it is denoted by $\{\mathcal{D}_{q,n}^{(j)}\}_{j=1}^{|\mathcal{D}_{q,n}|}$ (we denote $|\mathcal{A}|$ as the cardinality of set $\mathcal{A}$) for the $q^{th}$ node at the $n^{th}$ time instant, where $(j)$ denotes the $j^{th}$ entry of the dictionary. This dictionary is filled with tuples of innovations, whose contents denoted by $\{\mathcal{I}_{q,n}^{(j)}\}_{j=1}^{|\mathcal{D}_{q,n}|}$ and corresponding observations
$\{\mathcal{X}_{q,n}^{(j)}\}_{j=1}^{|\mathcal{D}_{q,n}|}$.

As we are dealing with non-linear estimation over a network, the local estimate of the implicit parameter $\Omega_{q}(n)$ for the $q^{th}$ node at the $n^{th}$ time instant in the RKHS, is given as follows:

\begin{equation}\label{rkhs1}
\Omega_{q}(n)=\eta\sum_{j=1}^{|\mathcal{D}_{q,n}|}\mathcal{I}_{q,n}^{(j)}<\phi(\mathcal{X}_{q,n}^{(j)}),.>_{\mathcal{H}}
\end{equation}
Fusing the observations by the matrix $C$ from the neighbouring nodes, we get the following modified current observation:
\begin{equation}
\textbf{x}^{'}_{q,n} = \sum_{\forall l}c_{lq}\textbf{x}_{l,n}
\end{equation}
Taking kernel inner product on both sides on (\ref{rkhs1}) with $\textbf{x}^{'}_{q,n}$, we arrive at the following adaptation:
\begin{equation}
y_{q,n}=\sum_{j=1}^{|\mathcal{D}_{q,n}|}\mathcal{I}_{q,n}^{(j)}<\phi(\mathcal{X}_{q,n}^{(j)}),\phi(\textbf{x}^{'}_{q,n})>_{\mathcal{H}}
\end{equation}
Note that we have invoked the identity for Hilbert adjoint operator, $<Tx,y>=<x,T^{*}y>$ where,
\begin{equation}
<\phi(\mathcal{X}_{q,n}^{(j)}),\phi(\textbf{x}^{'}_{q,n})>_{\mathcal{H}}=\exp(-\frac{\|\mathcal{X}_{q,n}^{(j)}-\textbf{x}^{'}_{q,n}\|^2}{\sigma^2})
\end{equation}
where $\sigma$ is the kernel spread parameter, a simulation parameter determined by Silverman's rule \cite{silverman1986density}, $x$ and $y$ are arbitrary elements from a Hilbert space endowed with  inner product $<\cdot,\cdot>$. $T$ is an operator and $T^{*}$ is its adjoint \cite{kreyszig1989introductory}. In our case, $T$ would be given by the Kronecker product of $C$ with an identity matrix.

The dictionary is updated based on an online vector quantisation approach similar to \cite{zhao2013fixed}. This controls the dictionary from growing unboundedly. The first proposed algorithm is formulated in Algorithm 1.

\begin{algorithm}
	\caption{ Quantised Diffusion KLMS (QDKLMS)}\label{qkmser}
	\begin{algorithmic}[1]
		\State \textbf{Initialise} step-size $\eta$, kernel width $\sigma$ and quantisation threshold $\epsilon>0$, and initial dictionary $\mathcal{D}_{0}= \{\delta_{0},x_{0}\}$ for all nodes in the network of network size $\mathcal{N}$

		\While{$|\mathcal{D}_{q,n}|\geq 1 \forall q$}
				\For {$q=1:|\mathcal{N}|$}
		\State $\textbf{x}^{'}_{q,n} = \sum_{\forall l \in \mathcal{N}_{q}}c_{lq}\textbf{x}_{q,n}$
		\State
		$y_{q,n}=\sum_{j=1}^{|\mathcal{D}_{q,n}|}\mathcal{I}_{q,n}^{(j)}<\phi(\mathcal{X}_{q,n}^{(j)}),\phi(\textbf{x}^{'}_{q,n})>_{\mathcal{H}}
		$
		\State $e_{q,n} =d_{q,n} - y_{q,n}$
		\State $e^{'}_{q,n}=\sum_{l \in \mathcal{N}_{q}} a_{ql}e_{l,n}$
		\State $j^{*}=\arg\min_{1\leq j\leq |\mathcal{D}_{q,n-1}|}
		\|\textbf{x}_{q,n}-\mathcal{D}_{q,n-1}^{(j)}\|$
		\If{$\|\textbf{x}_{q,n}-\mathcal{D}_{q,n-1}^{(j^{*})}\|\leq \epsilon$}
		\State $\mathcal{D}_{q,n}=\mathcal{D}_{q,n-1}$
		\State $\mathcal{I}_{q,n}^{(j^{*})}=\mathcal{I}_{q,n-1}^{(j^{*})}+\eta e_{q,n}^{'}$
		\Else
		\State $\mathcal{X}_{q,n}=\mathcal{X}_{q,n-1}\cup \textbf{x}_{q,n}, \mathcal{I}_{n}=\mathcal{I}_{n-1}\cup e_{q,k}^{'}$
		\EndIf
		\EndFor
		\EndWhile
		
	\end{algorithmic}
\end{algorithm}

The Algorithm 1 checks if the current observation is close to some observation in the dictionary. If yes, then the center of the corresponding observation in the dictionary is updated. Otherwise the new observation is appended to the dictionary.

Thus we can see from the above Algorithm 1 that the dictionary size is updated if the received observation at a given node $q$ is significantly ``different" (with respect to Euclidean norm) from all the observations in $q^{th}$ node's dictionary. This prevents the dictionary size at a particular node from growing unboundedly over the network. However, this algorithm does not give a way to prune the dictionary at all network nodes so as to discard ``unimportant" entries. The next algorithm, proposed in the following section provides us a method for online pruning the available dictionary, hence providing equivalent convergence with a much lesser storage requirement.

\section{Proposed Algorithm II : Fixed Budget Quantised Diffusion KLMS (FBQDKLMS)}
This section gives a method to prune the size of the available dictionary by techniques given in \cite{zhao2013fixed} which are based on online estimation of a term called ``significance". It is a measure of how much a particular entry of a dictionary $\mathcal{D}$ contributes to the overall learned hypothesis. Significance is estimated in the following manner depending on whether a new center is added, merged or pruned. In case a center is added, the significance, $E_{n}^{(j)}$, for $j^{th}$ entry of the dictionary at $n^{th}$ time instant and $q^{th}$ node is updated as follows \cite{zhao2013fixed}:
\begin{gather}\label{novelty}
E_{q,n}^{(j)}=\zeta E_{q,n-1}^{(j)}+\\\nonumber|e_{q,n}^{|\mathcal{D}_{q,n-1}|+1}|<\phi(\mathcal{D}_{q,n-1}^{(j)}),\phi(\mathcal{D}_{q,n}^{(|\mathcal{D}_{q,n-1}|+1)})>_{\mathcal{H}},\\ \nonumber \forall 1\leq j \leq |\mathcal{D}_{q,n-1}|
\end{gather}
where $\zeta$ is a forgetting factor such that $0<<\zeta\leq1$.
In case of merging, we update as follows:
\begin{gather}\label{merge1}
E_{q,n}^{(j\neq j^{*})}=\zeta E_{q,n-1}^{(j\neq j^{*})}+\\ \nonumber|\mathcal{I}_{q,n}^{(j\neq j^{*})}|<\phi(\mathcal{D}_{q,n-1}^{(j\neq j^{*})}),\phi(\mathcal{D}_{q,n}^{(j^{*})})>_{\mathcal{H}}
\end{gather}
$\lambda_{q,k}^{(j)}$ is a variable which is updated as:
\begin{gather}
\lambda_{q,n}^{(j)}=\zeta\lambda_{q,n-1}^{(j)}
\end{gather}
and,
\begin{gather}\label{merge2}
E_{q,n}^{(j^{*})}=\frac{|\mathcal{I}_{q,n}^{(j)}+\eta e_{q,n}|}{|\mathcal{I}_{q,n}^{(j)}|}\zeta E_{q,n-1}^{(j^{*})}+|\mathcal{I}_{q,n}^{(j)}+\\ \nonumber\eta e_{q,n}|<\phi(\mathcal{D}_{q,n-1}^{(j^{*})}),\phi(\mathcal{D}_{q,n}^{(j^{*})})>_{\mathcal{H}}
\end{gather}
In case of deletion/pruning of the $L^{th}$ dictionary entry,
\begin{gather}\label{prune}
E_{q,n}^{(j)}=E_{q,n-1}^{(j)}-|\mathcal{I}_{q,k}^{(j)}|\lambda_{q,n-1}^{(L)}<\phi(\mathcal{D}_{q,n-1}^{(j)}),\phi(\mathcal{D}_{q,n-1}^{(L)})>_{\mathcal{H}}
\end{gather}

\begin{gather}
\lambda_{n}^{(q,j)}=\zeta\lambda_{q,n-1}^{(j)}+1
\end{gather}
Using these distributed online estimates of significance, the proposed FBQDKLMS is given in Algorithm \ref{fbqkmser}.

\begin{algorithm}
	\caption{Fixed Budget Quantised Diffusion KLMS (FBQDKLMS)}\label{fbqkmser}
	\begin{algorithmic}[1]
			\State \textbf{Initialise} step-size $\eta$, kernel width $\sigma$ and quantisation threshold $\epsilon>0$, and initial dictionary $\mathcal{D}_{0}= \{\delta_{0},x_{0}\}$ for all nodes in the network of network $\mathcal{N}$

		\While{$|\mathcal{D}_{q,n}|\geq 1 \forall q$}
				\For {$q=1:|\mathcal{N}|$}
\State $\textbf{x}^{'}_{q,n} = \sum_{\forall l \in \mathcal{N}_{q}}c_{lq}\textbf{x}_{q,n}$
\State $y_{q,n}=\eta\sum_{j=1}^{|\mathcal{D}_{q,n-1}|}<\phi(\mathcal{D}_{q,n-1}^{(j)}),\phi(\textbf{x}^{'}_{q,n})>_{\mathcal{H}}\mathcal{I}_{q,n-1}^{(j)}$
\State $e_{q,n} =d_{q,n} - y_{q,n}$
		\State $e^{'}_{q,n}=\sum_{l \in \mathcal{N}_{q}} a_{ql}e_{l,n}$		
 \State $j^{*}=\arg\min_{1\leq j\leq |\mathcal{D}_{q,n-1}|}\|\textbf{x}_{q,n}-\mathcal{D}_{q,n-1}^{(j)}\|$
		\If{$\|\textbf{x}_{q,n}-\mathcal{D}_{n-1}^{(j^{*})}\|\leq \epsilon$}
		\State $\mathcal{D}_{q,n}=\mathcal{D}_{q,n-1}$
		\State $\mathcal{I}_{q,n}^{(j^{*})}=\mathcal{I}_{q,n-1}^{(j^{*})}+\eta e_{q,n}^{'}$
		\State Update significance $\{E_{q,n}^{(j)}\} \forall j$ as per eq. \State(\ref{merge1}) and eq. (\ref{merge2})
		\Else
		\State $\mathcal{D}_{q,n}=\mathcal{D}_{q,n-1}\cup \textbf{x}_{q,k}, \mathcal{I}_{q,n}=\mathcal{I}_{q,n-1}\cup e_{q,n}^{'}$
		\State Update significance $\{E_{q,n}^{(j)}\} \forall j$ as per eq. \State(\ref{merge1}) and eq. (\ref{merge2})
		\State Also update significance for newly added tuple as per
		\State eq. (\ref{novelty}).
		\EndIf
		\State $\mathcal{D}_{q,n-1}^{'}=\{ (I_{k}^{(j)},\textbf{x}_{k}^{(j)}) \in \mathcal{D}_{q,n-1}\forall j : E_{q,n}^{(j)}=\min\{E_{q,n}\}\}$
		\State $\mathcal{D}_{q,n}=\mathcal{D}_{q,n-1}-\mathcal{D}_{q,n-1}^{'}$
		\State Update significance $\{E_{q,n}^{(j)}\} \forall j$ as per eq. (\ref{prune})
		\EndFor
\EndWhile
	\end{algorithmic}
\end{algorithm}
In Algorithm 2 we have two independent measures of controlling the dictionary length. First, measure is the Euclidean proximity of an incoming observation to a member of dictionary. If there is a member in the dictionary which is in an $\epsilon$ neighborhood of the regressor, the member's contents are updated. Otherwise the new observation is added to the dictionary. On the other hand, a term called significance introduced in  \cite{zhao2013fixed}, which estimates the overall contribution of a member of the dictionary to the overall hypothesis learned, is estimated recursively. If the significance of a member is the least among all elements in the dictionary, then that particular member is deleted from the dictionary. Thus this makes the dictionary more flexible by giving mechanisms for both expanding and reducing a dictionary.
\section{Simulations}
\begin{figure}
	\includegraphics[width=0.35\textwidth]{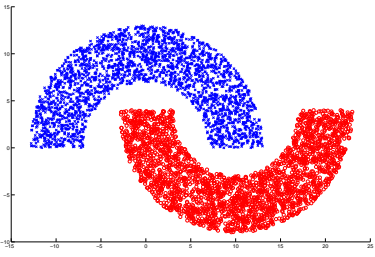}
	\caption{Scatter diagram for synthetic dataset b)}
	\label{fig:1}       
\end{figure}

\begin{figure}[!htbp]
	\includegraphics[width=0.35\textwidth]{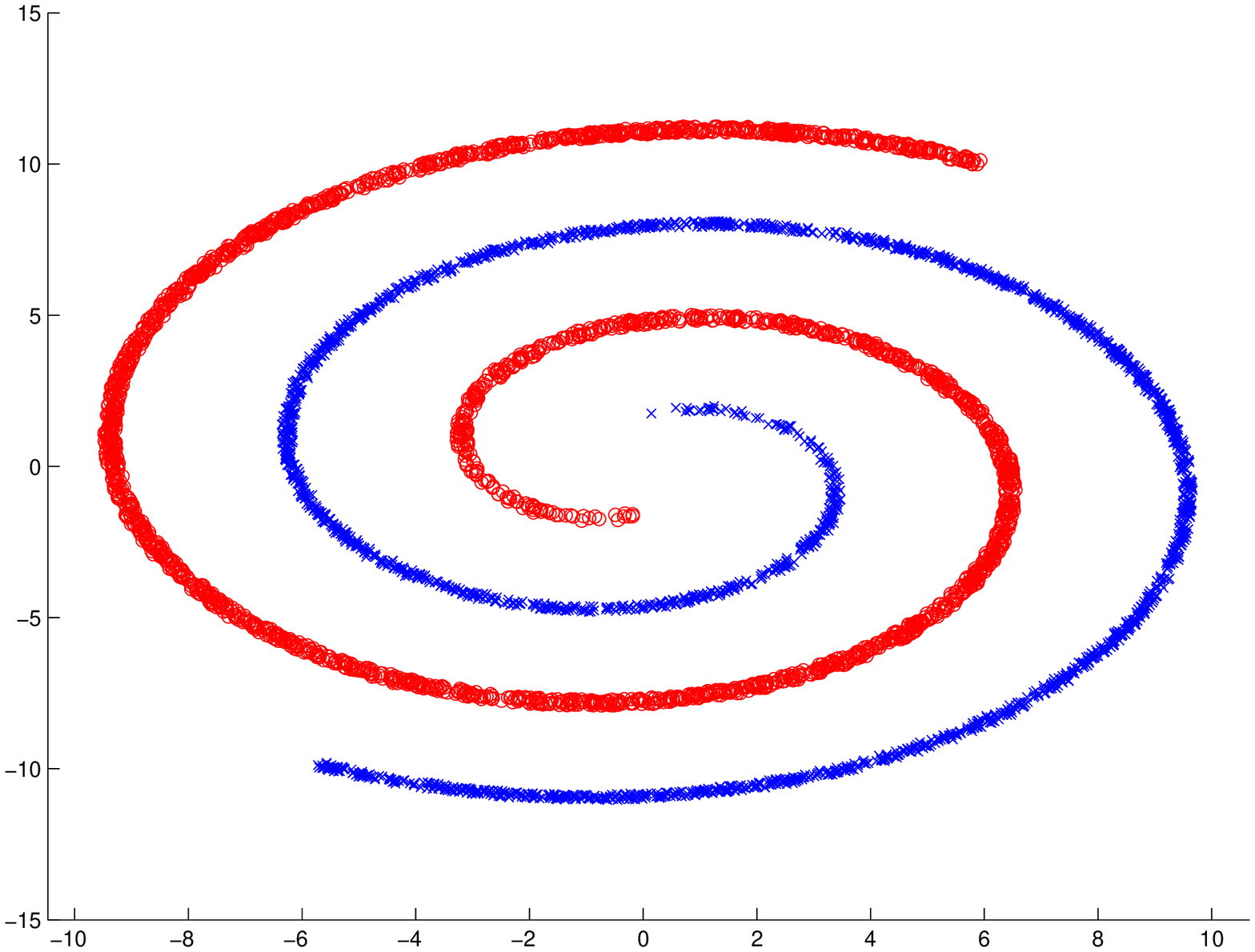}
	\caption{Scatter diagram for synthetic dataset c)}
	\label{fig:2}       
\end{figure}

In this section, we provide simulations to validate the two newly proposed algorithms against the existing literature.
We first discuss about the simulation setup used in this paper. We considered three simulation scenarios: a) the non-stationary channel from \cite{zhao2013fixed}, b) crescent moon dataset \cite{wild2008optimization} and c) spiral dataset \cite{haykin2004comprehensive}. The scatter diagrams of the crescent moon and spiral datasets are given Fig. \ref{fig:1} and Fig. \ref{fig:2}. For the proposed algorithms, network nodes  are considered which are assumed to produce data randomly from above mentioned considered datasets.

Subsequently, we discuss the simulation parameter values. Throughout, kernel width value, determined by Silverman's rule, was used and step-size $\eta$ is fixed to 0.1 to compare all the algorithms. The fixed budget was varied depending on the dataset such that both algorithms have similar steady-state dictionary size.

\begin{figure*}
	\includegraphics[width=0.85\textwidth,,height=5.5cm]{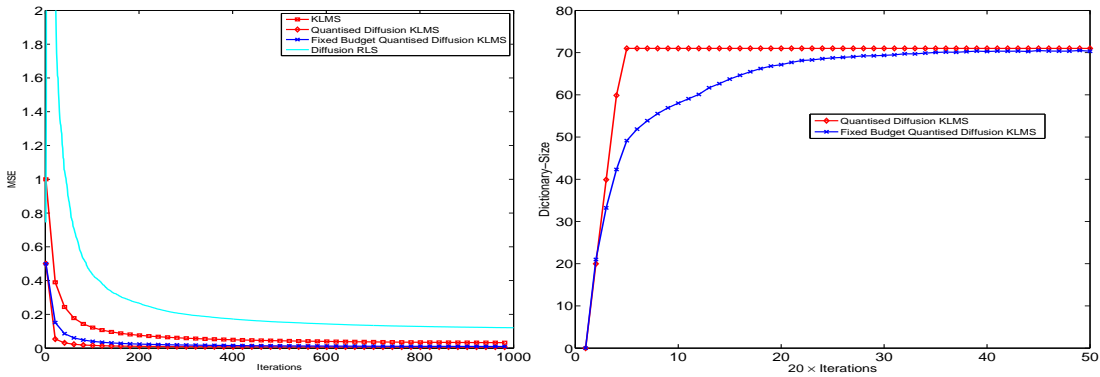}
	\caption{MSE and dictionary-size evolution for non-stationary channel}
	\label{fig:3}       
\end{figure*}

\begin{figure*}
	\includegraphics[width=0.85\textwidth,height=5.5cm]{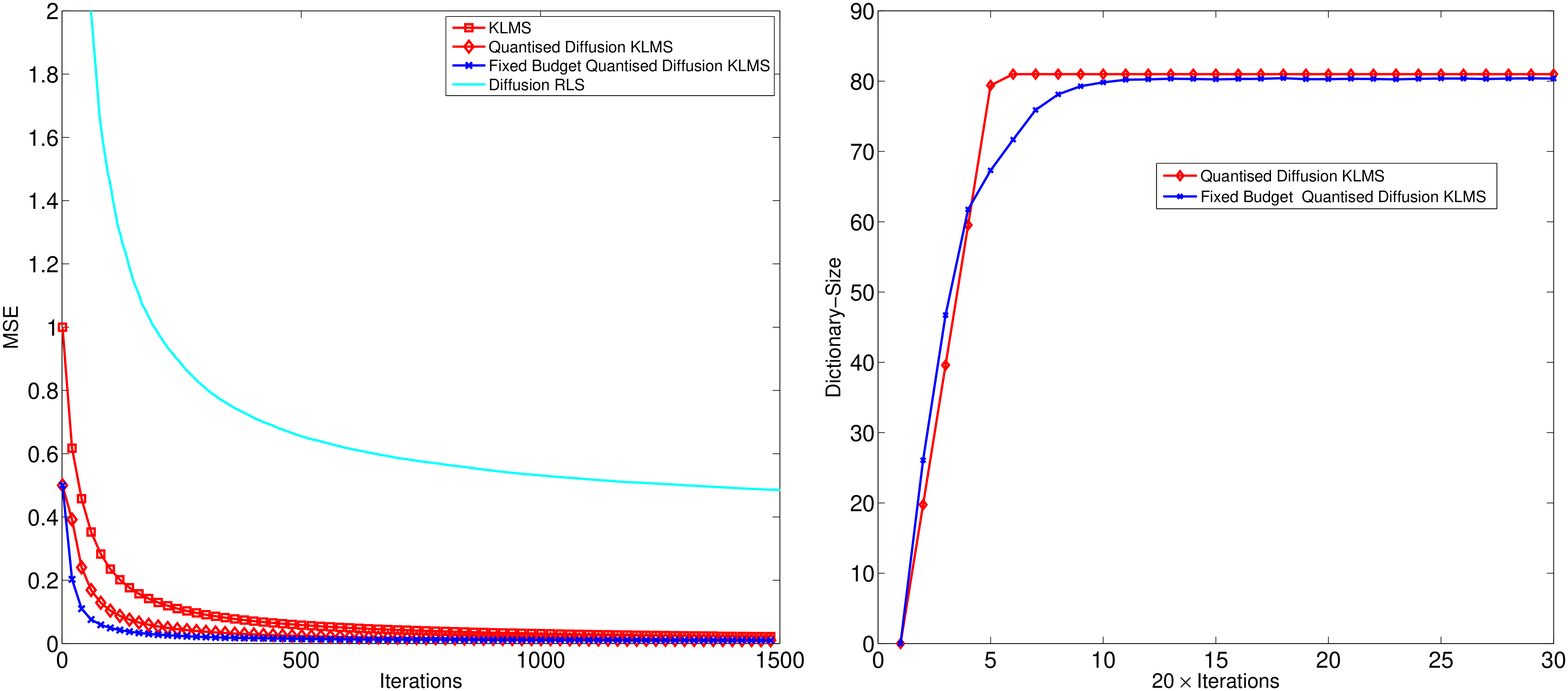}
	\caption{MSE and dictionary-size evolution for synthetic dataset b)}
	\label{fig:4}       
\end{figure*}

\begin{figure*}
	\includegraphics[width=0.85\textwidth,height=5.5cm]{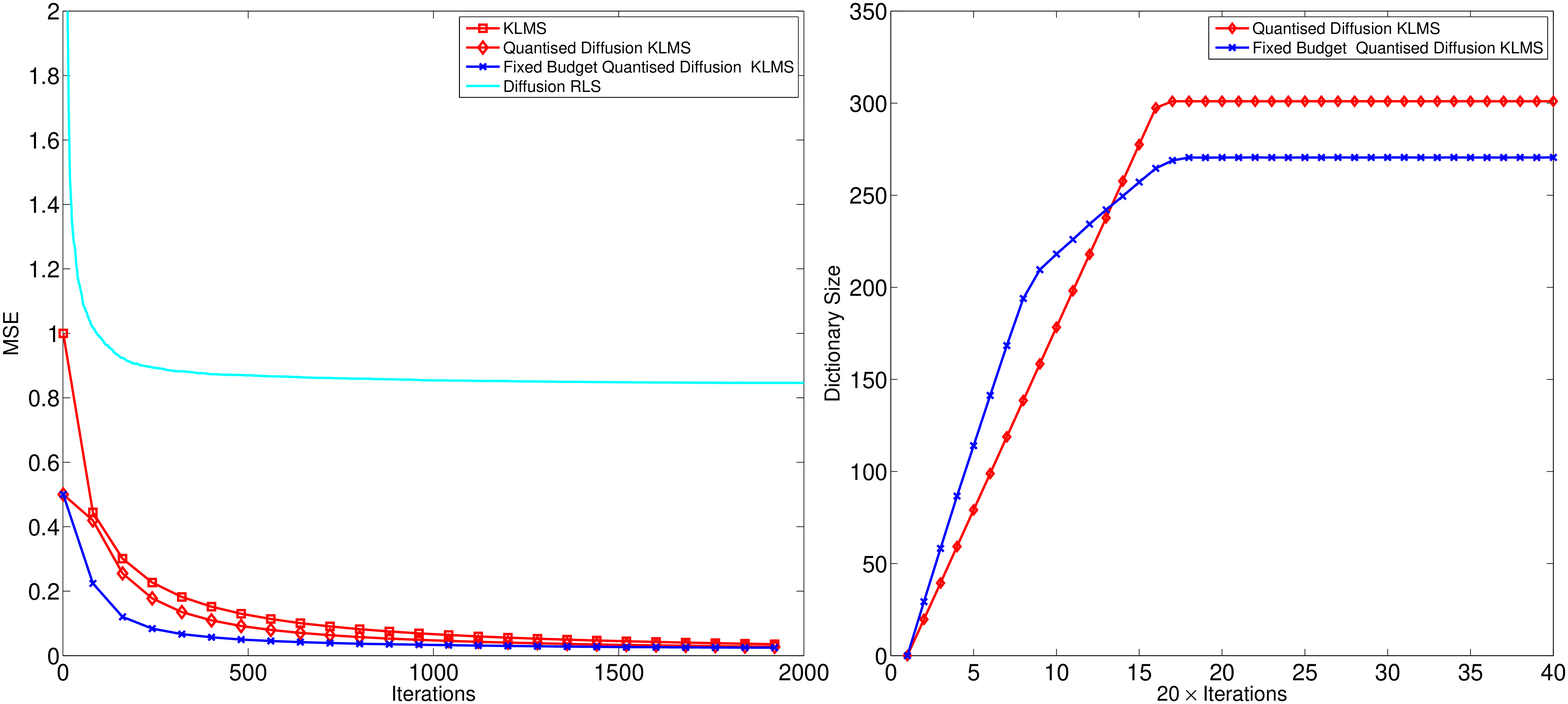}
	\caption{MSE and dictionary-size evolution for synthetic dataset c)}
	\label{fig:5}       
\end{figure*}

Finally, we provide the convergence results in Fig. \ref{fig:3}, Fig. \ref{fig:4} and Fig. \ref{fig:5}. In Fig. \ref{fig:3}, we consider the non-stationary equalization channel in \cite{zhao2013fixed} with binary input. It is observed that there is faster convergence in case of QDKLMS and FBQDKLMS as compared to KLMS with a single node. It is also observed that QDKLMS and FBQDKLMS exhibit similar transient behaviour with FBQDKLMS being tendentious to smaller dictionary-sizes. In Fig. \ref{fig:4} and Fig. \ref{fig:5}, we find that the proposed approaches QDKLMS and FBQDKLMS converge faster than KLMS, with FBQDKLMS having superior convergence as compared to QDKLMS with similar or lower storage requirement.

\begin{figure*}
	\includegraphics[width=0.85\textwidth]{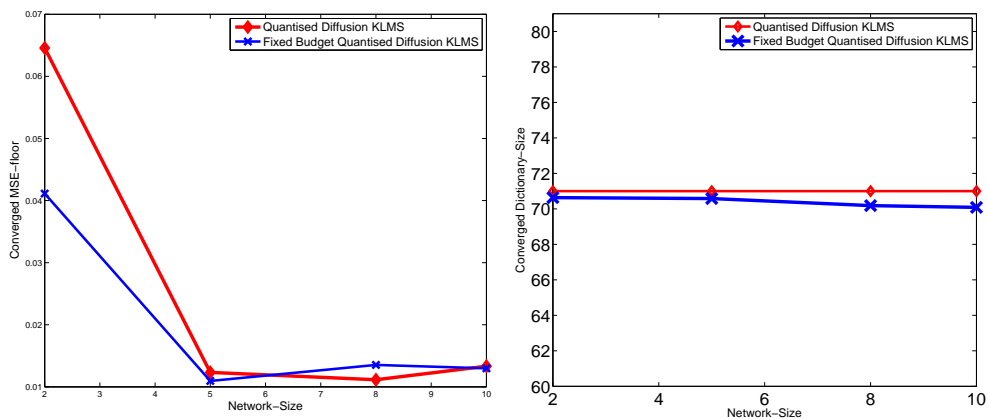}
	\caption{Variation of MSE floor and converged network-size with increase of number of nodes for non stationary equalisation.}
	\label{fig:6}       
\end{figure*}

\begin{figure*}
	\includegraphics[width=0.85\textwidth]{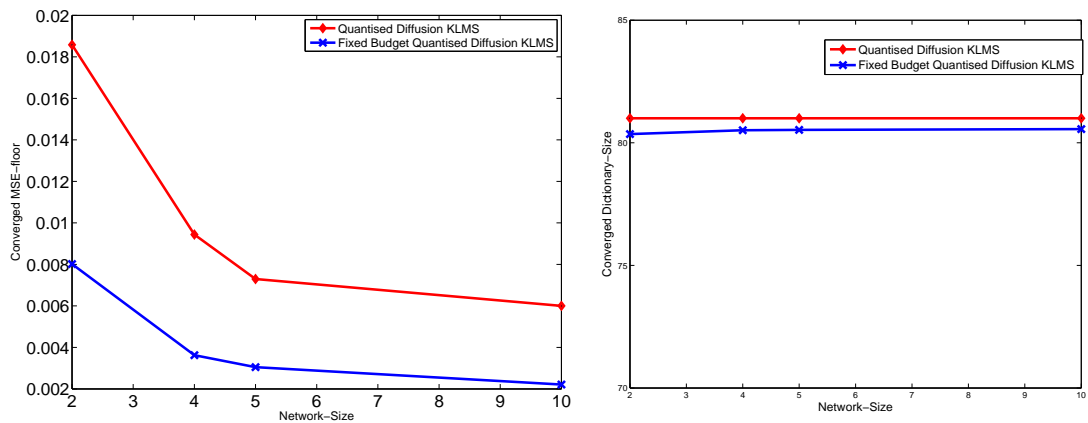}
	\caption{Variation of MSE floor and converged network-size with increase of number of nodes for crescent moon dataset b).}
	\label{fig:7}       
\end{figure*}

\begin{figure*}
	\includegraphics[width=0.85\textwidth]{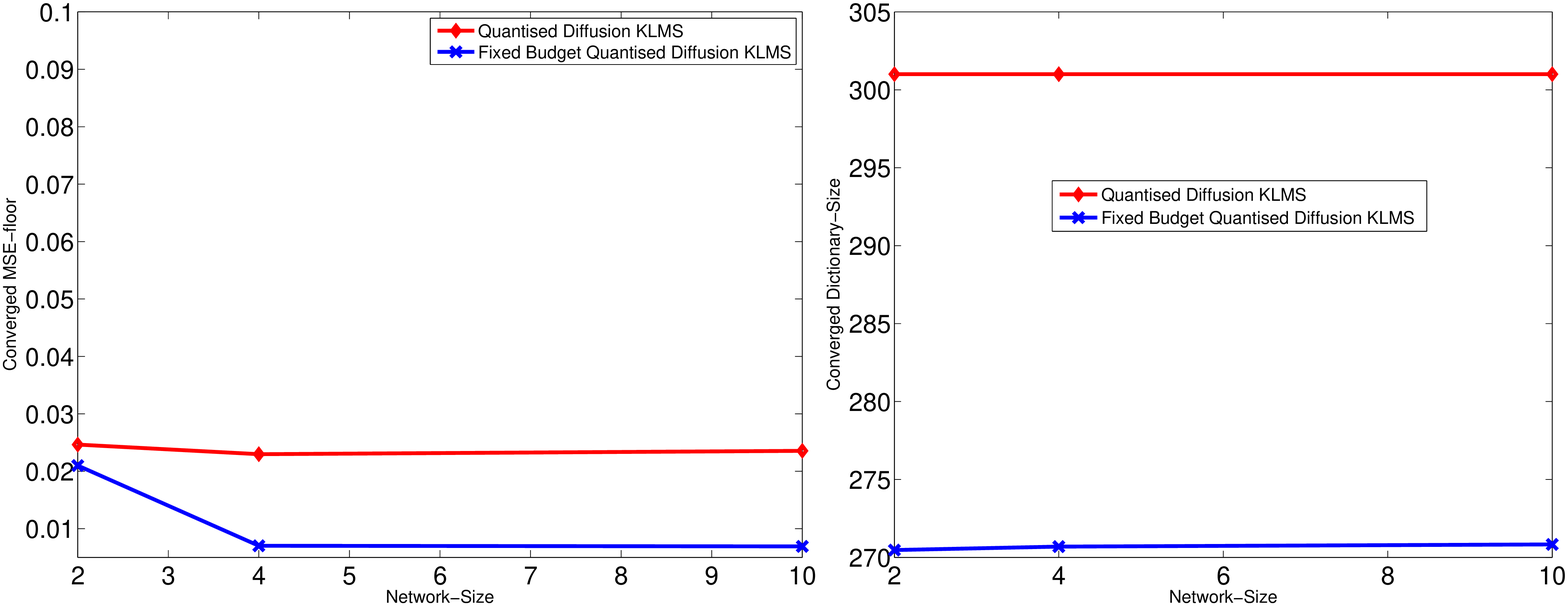}
	\caption{Variation of MSE floor and converged network-size with increase of number of nodes for spiral dataset c).}
	\label{fig:8}       
\end{figure*}

To see how the two proposed algorithms compare against each other whilst we scale the network-size we present simulations in Fig. \ref{fig:6}, Fig. \ref{fig:7} and Fig. \ref{fig:8}. All outputs of these simulations have been averaged over 200 monte-carlo iterations.

In Fig. \ref{fig:6}, we consider the non-stationary equalisation problem. We find that both the proposed algorithms exhibit similar decreasing trend of converged MSE floor as we scale the network size. We also see that the FBQDKLMS converges to a lower dictionary size requirement as compared to QDKLMS for smaller number of nodes.

In Fig. \ref{fig:7} and Fig. \ref{fig:8}, we compare the proposed algorithms on the crescent moon dataset. We find that the FBQDKLMS converges to a lower error floor as we scale the network size and converges to lower dictionary-size with respect to number of nodes in the inference network.

These simulations indicate that given a network with observations emanating for nodes equipped with finite storage, QDKLMS and FBQDKLMS algorithms are  robust and are also flexible for the assumed system models.

\section{Conclusion and Future Work}
In this paper two distributed kernel adaptive filtering algorithms, namely the quantised diffusion KLMS and the fixed budget quantised diffusion KLMS were introduced which work with a limited dictionary across all the network nodes. These algorithms have been found by simulations to be converging faster than other existing distributed adaptive filtering algorithms available in the literature. Also, the proposed algorithms need lesser memory and processing power requirement as they work with a finite dictionary at all nodes, and hence can be implemented in a practical system. Among the proposed algorithms, the FBQDKLMS is more preferred as its performance has been found to be comparable to QDKLMS and has been found to be tendentious to lower dictionary sizes. This work can be extended for applications like wireless sensor networks, distributed massive multiple input multiple output (MIMO) for 5G applications and distributed spectrum sensing in cognitive radio.


%





\bibliographystyle{spmpsci}      

\bibliography{paper}   

%
%

\end{document}